\documentclass[aps,preprint,epsfig,rotate]{revtex4}
\begin{document}
%\begin{doublespace}
\title{On the properties of the Uehling potential}

 \author{Alexei M. Frolov}
 \email[E--mail address: ]{afrolov@uwo.ca}

\affiliation{Department of Chemistry\\
 University of Western Ontario, London, Ontario N6H 5B7, Canada}

\date{\today}

\begin{abstract}

A number of properties of the Uehling potential are investigated. In particular, 
we determine the Fourier spatial resolution of the Uehling potential. The 
lowest-order correction on vacuum polarisation is re-written in terms of the 
electron density distribution function. We also discuss the consecutive 
approximations of the perturbation theory developed for the short-range Uehling
potential in the Coulomb few-body systems (atoms). The cusp problem is 
formulated for few-body systems in which particles interact with each other by 
the mixed (Coulomb + Uehling) potential. 

\end{abstract}

\maketitle
\newpage

\section{Introduction}

As is well known (see, e.g., \cite{AB}, \cite{Grei}) in the lowest order
approximation the effect of vacuum polarisation between two interacting 
electric charges is described by the Uehling potential $U(r)$ \cite{Uehl}. 
In \cite{FroWa1} we have derived the closed analytical formula for the 
Uehling potential. For atomic systems this formula is written in the 
following three-term form (in atomic units $\hbar = 1, m_e = 1, e = 1$)
\begin{eqnarray}
 U(2 b r) = \frac{2 \alpha Q}{3 \pi} \cdot \frac{1}{r} \Bigl[ 
 \int_1^{+\infty} exp(-2 \alpha^{-1} \xi r) \Bigl(1 + \frac{1}{2 \xi^2}
 \Bigr) \frac{\sqrt{\xi^2 - 1}}{\xi^2} d\xi \Bigr] \nonumber\\
 = \frac{2 \alpha Q}{3 \pi r} \Bigl[ \Bigl(1 + \frac{b^2
 r^2}{3}\Bigr) K_0(2 b r) - \frac{b r}{6} Ki_1(2 b r) - \Bigl(\frac{b^2
 r^2}{3} + \frac{5}{6}\Bigr) Ki_2(2 b r) \Bigr] \; \; \; , \label{Ueh}
\end{eqnarray}
where the notation $Q$ stands for the electric charge of the nucleus, $b = 
\alpha^{-1}$ and $\alpha = \frac{e^2}{\hbar c} \approx \frac{1}{137}$ is the 
dimensionless fine-structure constant. Here and below $\hbar = \frac{h}{2 
\pi}$ is the reduced Planck constant (also called the Dirac constant), $e$ 
is the electric charge of the positron and $m_e$ is the mass of the electron 
(= mass of the positron). 

In Eq.(\ref{Ueh}) $K_0(a)$ is the modified Bessel function of zero order 
(see, e.g, \cite{GR}), i.e.
\begin{eqnarray}
 K_0(z) = \int_0^{\infty} exp(-z \cosh t) dt = \sum_{k=0}^{\infty} 
 (\psi(k+1) + \ln 2 - \ln z) \frac{z^{2 k}}{2^{2 k} (k!)^2} \; \; \; , 
 \label{macd}
\nonumber
\end{eqnarray}
where $\psi(k)$ is the Euler $psi$-function defined by Eq.(8.362) from
\cite{GR}. The functions $Ki_1(z)$ and $Ki_2(z)$ in Eq.(\ref{Ueh}) are the
recursive integrals of the $K_0(z) \equiv K_0(z)$ function, i.e.
\begin{eqnarray}
 Ki_1(z) = \int_z^{\infty} Ki_0(z) dz \; \; \; , \; \; \; and \; \; \;
 Ki_n(z) = \int_z^{\infty} Ki_{n-1}(z) dz  \; \; \; , \label{repin}
\end{eqnarray}
where $n \ge 1$.

The closed analytical formula for the Uehling potential derived in \cite{FroWa1} 
allows one to obtain substantial acceleration of numerical algorithms used to
evaluate the effects of vacuum polarisation in various few-electron atoms and ions. 
On the other hand, the simple three-term formula, Eq.(\ref{Ueh}), can also be used 
to perform theoretical analysis of the vacuum polarisation itself. In this study 
we report some interesting results of our recent investigations. These results can 
be useful in numerical applications and may lead us to better understanding of 
vacuum polarisation in atomic systems. In particular, in this study we derive the 
general expression for the lowest-order correction on the vacuum polarisation in 
atoms and ions written with the use of the electron density distribution function 
$\rho(r)$. The arising expression has a number of advantages for theoretical 
analysis of vacuum polarisation in light atoms and ions. Then we determine the 
spatial Fourier resolution of the Uehling potential $U(r)$. This leads us to the 
formula for the spatial Fourier resolution of the actual electrostatic field which 
is the sum of the regular Coulomb potential and short-range Uehling potential.  

Another interesting problem is the application of the perturbation theory to 
determine various corrections produced by the short-range Uehling potential. It is 
shown in the fourth Section that all consecutive approximations produced by the
Uehling potential can be determined with the use of the representations of the 
O(2,1)-algebra(s). Such an algebra describes the dynamical representations of the
original Coulomb system (or atom). We also consider the cusp problem for few-body 
systems of interacting electrically charged particles.    

\section{Uehling correction and electron density distribution}

In atomic systems the lowest-order correction on vacuum polarization $(\Delta 
E)_{V.P.}$ to the total energy of the $LS-$bound state is written in the following 
general form
\begin{eqnarray}
 (\Delta E)_{V.P.} = \int_{0}^{\infty} \Psi^{*}_{LS}({\bf r}) U(r) \Psi_{LS}({\bf r}) 
 d^3{\bf r} = 4 \pi \int_{0}^{\infty} \Psi_{LS}(r) U(r) \Psi_{LS}(r) r^2 dr  
 \nonumber \\
 = 4 \pi \int_{0}^{\infty} U(r) \rho_{LS}(r) r^2 dr \label{eq1}
\end{eqnarray}
where $U(r)$ is the central Uehling potential Eq.(\ref{Ueh}). In Eq.(\ref{eq1}) the 
notation $\Psi_{LS}({\bf r})$ stands for the bound state wave function, while 
$\Psi_{LS}(r)$ means the radial part of the total wave function. The index $LS$ 
designates the angular momentum and total (electron) spin of the bound state. We 
assume that the bound state wave function $\Psi_{LS}(r)$ is know. Furthermore, such 
a function can always be chosen as a real function (or it can be transformed into 
the real function). For the real wave function one finds $\Psi^{*}_{LS}(r) = 
\Psi_{LS}(r)$. The function $\rho(r) = \Psi^{*}_{LS}(r) \cdot \Psi_{LS}(r) = 
\Psi_{LS}(r) \cdot \Psi_{LS}(r)$ is the electron density distribution in the bound 
$LS-$atomic state. The formula which contains the electron density function $\rho(r)$ 
can also be applied in those cases when the atomic system occupies one of its `mixed' 
states. 

The expression for the $(\Delta E)_{V.P.}$, Eq.(\ref{eq1}), can be transformed to 
the different form with the use of our formula from \cite{FroWa1}. Indeed, we can
write Eq.(\ref{eq1}) as the sum of the three following terms $T_1, T_2$ and $T_3$, 
i.e. 
\begin{eqnarray}
 (\Delta E)_{V.P.} = \frac{2 \alpha Q}{3 \pi} (T_1 + T_2 + T_3)
\end{eqnarray}
The explicit expressions for the $T_1, T_2$ and $T_3$ terms are reduced to the form 
\begin{eqnarray}
 T_1 = \frac{1}{4 b^2} \int_{0}^{\infty} \Bigl[ 1 + \frac{x^2}{12} \Bigr]
 K_0(x) \rho(\frac{1}{2 b} x) x dx \label{eq51} \\
 T_2 = -\frac{1}{48 b^2} \int_{0}^{\infty} Ki_1(x) \rho(\frac{1}{2 b} x) x^2 dx
 \label{eq52} \\
 T_3 = -\frac{1}{4 b^2} \int_{0}^{\infty} Ki_2(x) \Bigl( \frac{x^2}{12} + \frac56 
 \Bigr) \rho(\frac{1}{2 b} x) x dx \label{eq53}
\end{eqnarray}
where $x = 2 b r$. The formulas for the $T_2$ and $T_3$ terms can be simplified by using 
a few different methods, e.g., integration by parts. Finally, the formula for the $T_2$ 
term takes the form
\begin{eqnarray}
 T_2 = - \frac{1}{48 b^2} \int_{0}^{\infty} K_0(x) \Phi(x) dx = 
 - \frac{1}{48 b^2} \int_{0}^{\infty} K_0(x) \Bigl[ \int^{x}_0 \rho(\frac{1}{2 b} y) 
 y^2 dy \Bigr] dx \label{eq521} 
\end{eqnarray}
where the function $\Phi(x)$ is
\begin{eqnarray}
 \Phi(x) = \int^{x}_0 \rho(\frac{1}{2 b} y) y^2 dy \label{eq523} 
\end{eqnarray}
The term $T_3$ is transformed into a similar form
\begin{eqnarray}
 T_3 = - \frac{1}{48 b^2} \int_{0}^{\infty} K_0(x) \Phi_1(x) dx = 
 - \frac{1}{48 b^2} \int_{0}^{\infty} K_0(x) \Bigl[ \int^{x}_0 dz \int^{z}_0
 \Bigl( \frac{y^3}{12} + \frac56 y \Bigr) \rho(\frac{1}{2 b} y) dy \Bigr] dx 
 \label{eq525} 
\end{eqnarray}
where the function $\Phi_1(x)$ is
\begin{eqnarray}
 \Phi_1(x) = \int^{x}_0 dz \int^{z}_0 \Bigl[ \frac{y^3}{12} + \frac56 y \Bigr] 
 \rho(\frac{1}{2 b} y) dy \label{eq527} 
\end{eqnarray}

\section{Fourier spatial resolution of the Uehling potential}

In various problems known in nuclear, atomic and molecular physics and in Quantum 
Elecrodynamics it is often important to apply the correct analytical expression for the 
Fourier spectral resolution of the Uehling potential. It is clear $a$ $priori$ that the 
Uehling potential $U(r)$ is a static potential which also is a central potential. 
Therefore, we can expect that its Fourier spectral resolution is a superposition of plane 
waves of zero frequency. Moreover, it can be shown that all these plane waves are 
longitudinal, i.e. they are oriented along the spatial vector ${\bf k}$. The Fourier 
spatial resolution (or Fourier resolution, for short) of the Uehling potential 
is written in the form
\begin{eqnarray}
 U({\bf r}) = \int^{+\infty}_{-\infty} \int^{+\infty}_{-\infty} 
 \int^{+\infty}_{-\infty} exp(-\imath {\bf k} \cdot {\bf r}) u({{\bf k}})
 \frac{d^3{\bf k}}{(2 \pi)^3} \label{eqf1} 
\end{eqnarray}
where ${\bf k} = (k_x, k_y, k_z)$ is the wave vector and $u_{{\bf k}}(r)$ is the 
unknown spectral function. Our goal in this Section is to obtain the closed analytical 
formula for this spectral function and compare it with the spectral function of the pure 
Coulomb potential. 

From Eq.(\ref{eqf1}) one finds the following equation for the spectral function  
$u({{\bf k}}) = u(k)$
\begin{eqnarray}
 u(k) = \int \int \int U(2 b r) exp(\imath {\bf k} \cdot {\bf r}) d^3{\bf r} = 
 4 \pi \int^{+\infty}_0 j_0(k r) U(2 b r) r^2 dr \label{eqf2} 
\end{eqnarray}
where $k = \sqrt{k^2_x + k^2_y + k^2_z}$ is the radial component of the wave vector
and $j_0(x)$ is the spherical Bessel function of zero order. Note also that the 
integration in the last formula is performed over the area occupied by the electric 
charges and electric field. In Eq.(\ref{eqf2}) we have used the fact that the 
Uehling potential is spherically symmetric and, therefore, we can integrate over 
all angular variables. By using Eq.(\ref{Ueh}) and expression for the $j_0(x) = 
\frac{sin x}{x}$ function one finds the following formula for the spectral function 
$u(k)$
\begin{eqnarray}
 u(k) = \frac{2 \alpha^3 Q}{3} \int_{1}^{+\infty} \frac{1}{t^2 + a^2} \Bigl( 1 + 
 \frac{1}{2 t^2} \Bigr) \frac{\sqrt{t^2 - 1}}{t^2} dt \label{eqf3} 
\end{eqnarray}
where $a = \frac{k}{2 b} = \frac{k \alpha}{2}$. The integral in the last equation can
be determined with the use of the substitution $u = \frac{t}{\sqrt{t^2 - 1}}$. It 
reduces the integral to the form
\begin{eqnarray}
 I = \int_{1}^{+\infty} \frac{1}{t^2 + a^2} \Bigl( 1 + \frac{1}{2 t^2} \Bigr) 
 \frac{\sqrt{t^2 - 1}}{t^2} dt = -\frac12 \int_1^{+\infty} \frac{1}{u^2 (1 + 
 a^2) - a^2} \cdot \Bigl( 3 - \frac{1}{u^2} \Bigr) \frac{du}{u^2} \label{eqf4} 
\end{eqnarray}
Then by using the substitution $y = \frac{1}{u}$ one transforms this integral to the
form
\begin{eqnarray}
 I = \frac{1}{2 a^2} \int_{0}^{1} \frac{(3 - y^2) y^2 dy}{c^2 - y^2} = \frac{1}{2 a^2}
 \Bigl[ -\frac53 + \frac{1}{a^2} - \frac12 \frac{\sqrt{a^2 + 1}}{a} \Bigl(2 - 
 \frac{1}{a^2}\Bigr) ln \Bigl(\frac{\sqrt{1 + a^2} + a}{\sqrt{1 + a^2} - a}\Bigr) \Bigr]
 \label{eqf5} 
\end{eqnarray}
where $c^2 = \frac{a^2 + 1}{a^2}$. 

The final formula for the (spatial) spectral function $u(k)$ takes the form
\begin{eqnarray}
 u(k) =  \frac{\alpha^3 Q}{3 a^2} \Bigl[-\frac{5}{3} + \frac{1}{a^2} - 
 \frac12 \frac{\sqrt{a^2 + 1}}{a} \Bigl(2 - \frac{1}{a^2}\Bigr) ln \Bigl(\frac{\sqrt{1 + 
 a^2} + a}{\sqrt{1 + a^2} - a}\Bigr) \Bigr] \label{eqf6} 
\end{eqnarray}
or, since $a = \frac{k \alpha}{2}$
\begin{eqnarray}
 u(k) =  \frac{8 \alpha Q}{3 k^2} \Bigl[-\frac{5}{6} + \frac{2}{\alpha^2 k^2} - 
 \frac{\sqrt{\alpha^2 k^2 + 4}}{\alpha k} \Bigl(1 - \frac{2}{\alpha^2 k^2}\Bigr) 
 ln \Bigl(\frac{\sqrt{4 + \alpha^2 k^2} + k \alpha}{\sqrt{4 + \alpha^2 k^2} - k \alpha}\Bigr)
 \Bigr] \label{eqf65} 
\end{eqnarray}
As follows from the last formula the first term in the expression for $u(k)$ is $\sim k^{-2}$. 
This gives the asymptotic of the spectral function of the Uehling potential at large wave 
numbers $k$. The spectral function for the Coulomb potential between two interacting 
electrically charged particles is $u_C(k) = \frac{4 \pi}{k^2}$ (in atomic units) (see, e.g., 
\cite{LLE}). There are also a number of differences between the two spectral functions $u(k)$, 
Eq.(\ref{eqf65}), and $u_C(r)$. First, the $u(k)$ function contains various powers of the 
inverse wave number $k^{-1}$, while $u_C(k)$ is an exact quadratic function. Second, the $u(k)$ 
function, Eq.(\ref{eqf65}), contains inverse powers of the fine-structure constant $\alpha$, 
while the $u_C(k)$ function does not depend upon $\alpha$, if its expressed in atomic units. 
These two differences can also be found for the corresponding electric fields which are 
determined as the gradients of the corresponding potentials. The ${\bf k}$-component (Fourier 
component) of the total electric filed (Coulomb + Uehling potentials) is ${\bf E}_{{\bf k}} = 
- \imath \frac{4 \pi}{k^2} (1 + W(k)) \cdot {\bf k}$, where the function $W(k)$ is
\begin{eqnarray}
 W(k) =  \frac{2 \alpha Q}{3 \pi} \Bigl[-\frac{5}{6} + \frac{2}{\alpha^2 k^2} - 
 \frac{\sqrt{\alpha^2 k^2 + 4}}{\alpha k} \Bigl(1 - \frac{2}{\alpha^2 k^2}\Bigr)
 ln \Bigl(\frac{\sqrt{4 + \alpha^2 k^2} + k \alpha}{\sqrt{4 + \alpha^2 k^2} - k \alpha}\Bigr) 
 \Bigr] \label{eqf7} 
\end{eqnarray}
As follows from this formula each of the ${\bf E}_{{\bf k}}$ components is oriented along 
the wave vector ${\bf k}$, i.e. the total electric field contains only the longitudinal 
components (there are no non-zero transverse components) and these components do not depend 
upon the time $t$. Also, as one can see from the formula Eq.(\ref{eqf7}) in the lowest-order
approximation the contribution from the Uehling potential can be described as a small change
of the nuclear electric charge $Q \rightarrow Q \Bigl( 1 - \frac{5 \alpha}{9 \pi} \Bigr)$.  

This result indicates clearly that the Uehling potential (i.e. the potential which is 
responsible for the lowest order correction on vacuum polarization) can easily be 
incorporated into the basic equations of Quantum Electrodynamics from the very beginning 
(see, e.g, \cite{AB} and \cite{Heit}). Briefly, the total (i.e. Coulomb + Uehling) 
interparticle potential must be obtained during summation of the longitudinal and scalar 
components of the electromagnetic field. Note that for the first time the formulas 
Eq.(\ref{Ueh}) and Eq.(\ref{eqf6}) were produced by Pauli and Rose in \cite{PauliRose} (see 
the formula in footnote 4 and the very last equaltioni in that paper). 

\section{Consecutive approximations upon the fine structure constant}

In general, the short-range Uehling potential can be considered as a small correction to
the Coulomb interparticle potential. The contributions of the Uehling potential can be 
taking into account with the use of the perturbation theory. The small parameter in this
theory is the fine structure constant $\alpha$ (or the factor 2 $\alpha$). Based on the
existence of the radial O(2,1)-algebra for the original Coulomb two-body system (atom) we
can find a very transparent physical sense for each order of approximation used in this 
procedure. Below in this Section, we restrict ourselves to the analysis of the Coulomb 
two-body systems. The cases of the Coulomb three- and four-body systems can be considered 
analogously. The Coulomb two-body system (or atom) includes the heavy nucleus with the 
electric charge $Qe$ and electron with mass $m_e$ and electric charge $-e$. The 
corresponding Schr\"{o}dinger equation (in atomic units) takes the following form 
\begin{eqnarray}
 \hat{H} \Psi = \Bigl[ -\frac12 \Bigl( \frac{\partial^2}{\partial r^2} + \frac{2}{r} 
 \frac{\partial}{\partial r} \Bigr) + \frac{\hat{L}^2}{r^2} - \frac{Q}{r} \Bigr] \Psi 
 = E \Psi \label{Eq1} 
\end{eqnarray}
where $\hat{H}$ is the Hamiltonian, $E$ is the energy of the system, $\hat{L^2}$ is the 
operator of the angular momentum of the system (atom) and $\Psi$ is the wave function. 
The operator $\hat{L^2}$ depends only upon two angular variables ($\Theta$ and $\phi$) 
and it commutes with any operator which depends upon the radial variable $r$. The 
explicit form of the $\hat{L^2}$ operator is
\begin{eqnarray}
 \hat{L^2} = - \Bigl[ \frac{1}{\sin\Theta} \frac{\partial}{\partial \Theta} \Bigl(
 sin\Theta \frac{\partial}{\partial \Theta} \Bigr) + \frac{1}{\sin^2\Theta}
 \frac{\partial^2}{\partial \phi^2} \Bigr]
\end{eqnarray}
The eigenfunctions of the $\hat{L^2}$ operator are the spherical harmonics (see, e.g., 
\cite{AB}) $Y_{\ell m}(\Theta,\phi)$ and $\hat{L^2} Y_{\ell m}(\Theta,\phi) = \ell 
(\ell + 1) Y_{\ell m}(\Theta,\phi)$.
  
In \cite{Bar} (see also \cite{Fro821} and \cite{Fro85}) it was shown that the three 
following operators
\begin{eqnarray}
  S = \frac12 r \Bigl( p^2_r + \frac{\hat{L}^2}{r^2} + 1 \Bigr) \; \; \; , \; \; \;
  T = r p_r \; \; \; \; \; and \; \; \; \; \;
  U = \frac12 r \Bigl( p^2_r + \frac{\hat{L}^2}{r^2} - 1 \Bigr) \label{Eq3}
\end{eqnarray}
form the non-compact O(2,1)-algebra with the commutation relations
\begin{eqnarray}
  [ S, T ] = -\imath U \; \; \; , \; \; \; [ T, U ] = \imath S \; \; \; , \; \; \;
  [ U, S ] = -\imath T \label{Eq4}
\end{eqnarray}
The Casimir operator of this algebra $C_2 = S^2 - U^2 - T^2$ equals to the $\hat{L}^2$ 
operator of the angular momentum. The exact coincidence of the Casimir operator $C_2$ 
of the radial algebra $O(2,1)$ with the Casimir operator $C_2 = L^2$ of the algebra of
three-dimensional rotations $O(3)$ is related with the complementarity of the 
corresponding representations in the space of more general $Sp(6 N,R)$-algebra 
representation (so-called Moshinsky complementarity of representations, see, e.g., 
\cite{Mos}).  

Based on this fact we can transform the operator $r (\hat{H} - E) = [(S + U) - E (S - U) 
+ Q]$ from the Schr\"{o}dinger equation by applying the unitary transformation $exp(\imath 
\beta T)$, where $\beta$ is the real parameter and $T$ is the generator from 
Eq.(\ref{Eq3}). Indeed, by using the well known Hausdorff formula
\begin{eqnarray}
  exp(-\imath \beta T) (S \pm U) exp(\imath \beta T) = exp(\pm \beta T) (S \pm U) 
  \label{Eq5}
\end{eqnarray}
Now, we can chose $\beta = ln\sqrt{-2 E}$ and reduce the original Schr\"{o}dinger equation
to the form
\begin{eqnarray}
  [ \sqrt{- 2 E} S - Q ] \phi(r,\Theta,\phi) = 0 \label{Eq6}
\end{eqnarray}
The operator $S$ has the discrete spectrum \cite{Bar}, \cite{Kog} which is written in the 
general form $S \mid n, \ell, m \rangle = n  \mid n, \ell, m \rangle$, where $n = 1, 2, 3, 
\ldots$ is an integer positive number. From Eq.(\ref{Eq6}) one finds $E_n = -\frac{Q^2}{2 
n^2}$ (the energy spectrum in atomic units) and $\mid n, \ell, m \rangle = A \phi(r,
\Theta, \phi)$, where $A$ is an arbitrary numerical constant. All these facts and equations 
are very well known for the Coulomb two-body systems (atoms). It appears that we can modify 
our procedure for pure Coulomb systems to include the Uehling potential. Such a modification 
is discussed below.

First, we note that from Eq.(\ref{Eq3}) it follows that $r = S + U$. Second, the modified
Uehling potential $r U(r)$ is written in the form:
\begin{eqnarray}
 r U(2 b r) = \frac{2 \alpha Q}{3 \pi} \Bigl[ \Bigl(1 + \frac{b^2
 r^2}{3}\Bigr) K_0(2 b r) - \frac{b r}{6} Ki_1(2 b r) - \Bigl(\frac{b^2
 r^2}{3} + \frac{5}{6}\Bigr) Ki_2(2 b r) \Bigr] \; \; \; , \label{UehX}
\end{eqnarray}
This means that the Uehling potential is a function of the $2 b r = \frac{2}{\alpha} (S + 
U)$ variable. Such a variable contain the small dimensionless parameter (the fine structure 
constant) $\alpha$. In Eq.(\ref{UehX}) the parameter $\frac{2}{\alpha}$ plays the role of 
the cut-off parameter, since all modified Bessel functions $K_0(2 b r), Ki_1(2 b r)$ and 
$Ki_2(2 b r)$ decrease exponentially at large $r$. Briefly, this means that we can restrict
the power series of the $r U(2 b r)$ potential to a very few first terms and expectation
values of all these terms can be determined with the use of the commutation relations for the
O(2,1)-algebra, Eq.(\ref{Eq4}). In particular, for me it was interesting to know that in the 
lowest order approximation the effect of vacuum polarisation can be represented as a small 
change of the electric charge (screening).  

\section{Cusp problem}

As is well known in any Coulomb few-body system the expectation value of the following 
operator (in atomic units)
\begin{eqnarray}
 \hat{\nu_{ij}} = \frac{1}{\langle \delta({\bf r}_{ij}) \rangle}
 \delta({\bf r}_{ij}) \frac{\partial}{\partial r_{ij}} \label{eqc1}
\end{eqnarray}
equals to the product of the corresponding electric charges $q_1$ and $q_2$. In particular,
for the electron-nuclear Coulomb interaction written in atomic units te expectation value
of the $\hat{\nu}_{eN}$ operator equals $Q$ exactly. The operator, Eq.(\ref{eqc1}), has a 
great interest for the general theory of bound states in Coulomb few-body systems. On the
other hand, the coincidence of the expectation value of this operator with its expected 
value, e.g., with $Q$, is often used in numerical computations to test the quality of trial
wave functions. 

Now, consider the case when Uehling potential is added to the Coulomb potential directly 
into the Schr\"{o}dinger equation. This will lead to the change of the cusp condition between
each pair of interacting particles. The fundamental problem is to find the $\nu_{ij} = 
\langle \hat{\nu}_{ij} \rangle$ value for few-body systems interacting by the potential which
is the sum of Coulomb + Uehling potential. Formally, the answer can be written in the 
following form (in atomic units)
\begin{eqnarray}
 \nu_{ij} =  q_i q_j \Bigl\{ 1 - \frac{\alpha}{3 \pi} \bigl[ \frac53 + 2 \gamma - 
 2 ln \alpha + 2 ln r \Bigr] \Bigr\}  \label{eqc2}
\end{eqnarray}
where $\alpha$ is the fine structure constant, $\gamma \approx$ 0.577215$\ldots$ is the 
Euler's constant. Note that the explicit expression for the $\nu_{ij}$ expectation value, 
Eq.(\ref{eqc2}), is $r-$dependent and its finite limit at $r \rightarrow 0$ does not exist. 
On the other hand, the correct definition of the cusp-value includes two-particle 
delta-function. For non-relativistic systems this delta-function is a constant at the 
distances shorter than $\Lambda_e = \alpha a_0 = \alpha$ (in atomic units), where 
$\Lambda_e$ is the Compton wavelength. In general, the non-relativistic wave function 
cannot produce the actual electron density distribution at distances shorter than 
$\Lambda_e$. Instead it always gives the constant. Briefly, this means that we have to
assume that $r = C \alpha a_0 = C \alpha$ in Eq.(\ref{eqc2}), where $C$ is a numerical 
constant (close to unity) which is niformly related to the electron-nuclear (and 
electron-electron) delta-function. Therefore, the cusp expectation value (or $\nu_{ij}$ 
value) is finite and its numerical value is
\begin{eqnarray}
 \nu_{ij} =  q_i q_j \Bigl\{ 1 - \frac{\alpha}{3 \pi} \bigl[ \frac53 + 2 \gamma  
 + 2 ln C \Bigr] \Bigr\} \label{eqc3}
\end{eqnarray}
In other words, the Uehling potential produces a negative correction to the pure Coulomb 
cusp which is $\sim \alpha$. It will be very interesting to perform highly accurate 
computations for some few-body systems with the mixed interaction potential between 
electrically particles (Coulomb + Uehling). Then we can determine the expectation values 
of the actual electron-nuclear and electron-electron cusp values. Note that in modern 
highly accurate computations the cusp values are determined to the accuracy $\approx 1 
\cdot 10^{-10} - 1 \cdot 10^{-12}$ $a.u.$ The expected Uehling correction to the pure 
Coulomb cusp is $\approx 1 \cdot 10^{-3}$ (or 0.1 \%) and it can easily be detected in 
such computations. 
 
\section{Conclusion}

We have considered a number of properties of the Uehling potential. In particular, we
have derived the expression for the lowest-order correction to the vacuum polarisation
which contains the electron density distribution function $\rho(x)$ in atoms and/or in
molecules. Such a representation has a number of advantages in various applications, 
including actual calculations of the vacuum polarisation in the mixed atomic states 
and variational procedures used to improve the electron density distribution function 
$\rho(x)$ from `exact' QED-based Hamiltonians. We also derive the explicit expression 
for the Fourier spatial resolution of the Uehling potential. It is shown that such a 
resolution contains only longitudinal waves of zero frequency. The formula for the 
Fourier spatial resolution of the Uehling potential is investigated. Small corrections
produced by the short-range Uehling potential for actual Coulomb systems can be taking 
into account by a method based on the use of the non-trivial O(2,1)-group for the
original Coulomb few-body systems. In conclusion, we wish to note that the Uehling 
potential has many other unique properties which were never investigated in earlier 
works, e.g., the `corrected' cusp between two interacting electric charges. Some of 
these properties will be discussed in our future studies.


\begin{thebibliography}{10}

\bibitem{AB}A.I. Akhiezer and V.B. Beresteskii, {\it Quantum Electrodynamics}, (4th 
Ed., Nauka (Science), Moscow (1981)), Chps. 4 and 5 (in Russian).

\bibitem{Grei}W. Greiner and J. Reinhardt, {\it Quantum Electrodynamics}
(4th. Ed., Springer Verlag, Berlin, (2010)).

\bibitem{Uehl}E.A. Uehling, Phys. Rev. {\bf 48}, 55 (1935).

\bibitem{FroWa1}A.M. Frolov and D.M. Wardlaw, Eur. Phys. Jour. B {\bf 63}, 
339 - 350 (2012).

\bibitem{GR}I.S. Gradstein and I.M. Ryzhik, {\it Tables of Integrals, Series
and Products}, (5th ed., Academic Press, New York, (1994)).

\bibitem{LLE}L.D. Landau and E.M. Lifshitz, {\it The Classical Theory of 
Fields}, (4th Ed., Pergamon Press, New York (1979)), Chp. 6.

\bibitem{Heit}W. Heitler, {\it The Quantum Theory of radiation}, (3rd. ed., Oxford, 
UK (1954)).

\bibitem{PauliRose}W. Pauli and Rose, Phys. Rev. {\bf 49}, 462 (1936).

\bibitem{Bar}A.O. Barut and R. Raczka, {\it Theory of Group Representations and 
Applications}, (Warsaw: PNW (1977)), Chps. 11 and 12.

\bibitem{Fro821}A.M. Frolov, {\it Analytical solution of the many-body 
Shr\"{o}dinger equation}, Preprint IAE-3644/1, 17 p. (1982) (in Russian), 
unpublished.

\bibitem{Fro85}A.M. Frolov, J. Phys. B {\bf 23}, 2401 (1990).

\bibitem{Mos}M. Moshinsky and C. Quesne, J. Math. Phys. {\bf 11}, 1631 (1970).

\bibitem{Kog}A. Perelomov, {\it Generalized Coherent States and Their Applications}, 
(Springer Verlag, Berlin (1986)).

\end{thebibliography}
\end{document}